# SWOLLEN MICELLES PLUS HYDROPHOBICALLY MODIFIED HYDROSOLUBLE POLYMERS IN AQUEOUS SOLUTIONS: DECORATION VERSUS BRIDGING. A SMALL ANGLE NEUTRON SCATTERING STUDY.

Mohammed FILALI[+ +], Raymond AZNAR[+], Mattias SVENSON[+],

Grégoire PORTE[+], Jacqueline APPELL[+]*

+ Groupe de Dynamique des Phases Condensées, UMR5581 CNRS - Université Montpellier II,
C.C. 26 , 34095 Montpellier cedex , France
++ Laboratoire de Physique du Solide , Faculté des Sciences, Dharmehraz, BP1796 Atlas, FES, Morocco

**ABSTRACT:**

In this paper we examine the effective interactions introduced between the droplets of an oil in water microemulsion upon progressive addition of hydrophobically modified water soluble poly(ethylene oxide)-PEO using essentially small angle neutron scattering. To discuss the relative importance of decoration and bridging of the droplets we compare analogous samples with addition of a PEO grafted at both extremities with hydrophobic $C_{12}H_{25}$ chains (PEO-2m) or addition of a PEO grafted at one extremity only with a $C_{12}H_{25}$ chain (PEO-m). PEO-m or PEO-2m adsorb onto the droplets via their hydrophobic extremities and the droplets are found to retain their form and size upon addition of up to 40 hydrophobic $C_{12}H_{25}$ chains per droplet. When the volume fraction of droplets is less than about 10%, the effective interactions introduced by PEO-m or PEO-2m are found to be very different: PEO-m introduces a repulsive interaction while PEO-2m introduces an effective attractive interaction. This attractive interaction leads to an associative phase separation in the range of low volume fraction when a sufficient amount of PEO-2m is added. At higher volume fraction the contribution of PEO-m or PEO-2m to the interaction is a small repulsive contribution. We discuss qualitatively the experimental results in terms of decoration or bridging of the droplets by the PEO polymeric chains.

*To whom correspondence should be addressed - appell@gdpc.univ-montp2.fr

**INTRODUCTION**

The properties of complex or structured fluids are currently the subjects of many studies. In a number of cases, these properties are described using the picture of a transient network characteristic of physical gels [1]. We describe here a model system allowing for the study of such transient networks and discuss its static properties: structure and interactions. These are the basis of our forthcoming studies on the dynamical properties of such systems.

In the recent years other systems, parent to ours, have been studied [2-,6]. Among those aqueous solutions of hydrophobically modified water-soluble polymers have been largely investigated. They have been found to aggregate spontaneously by self-assembly of their hydrophobic parts. An important enhancement of the low shear viscosity of their solutions as concentration increases was thus observed. The earlier studies examined commercial polymers but investigations on well characterized samples was a pre-requisite to a proper description of the structure and properties of these solutions. Different groups synthesized well-characterized polymers (i.e. both as regard their mass and polydispersity as well as their degree of substitution) and studied their properties in solution, see e.g. [2,4,7-,14]. In many applications, associative polymers are found mixed with other constituents such as surfactants or latexes (see their use in water based surface coatings [2,8]), the next step has thus been naturally to study solutions of increasingly more complicated composition. Interactions between associative polymers and surfactants have been explored [15-16,23].

Aqueous solutions of the model associative polymers, where an hydrophobic alkyl chain is grafted at the two extremities of a poly(ethylene oxide), have been studied by different experimental techniques [4, 13, 14]. Above a critical concentration: "CAC" they self-assemble via their hydrophobic end chains forming small hydrophobic domains made of a small number of chains ( 5 to 25 chains depending both on the length of the PEO chain and the length of the hydrophobic chains). Bridging by the PEO chains of the hydrophobic domains can occur so that, by increasing concentration, larger and larger aggregates are formed and ultimately an interconnected network spans over the sample.

Turning to the properties of mixed solutions of associative polymer and surfactant, the point of view often adopted is to examine the perturbations introduced, upon addition of the surfactant, to the solution of associative polymers [15,17,18,19,20]. More recently we adopted another starting-point [21] where the end-grafted associative polymer and the non-ionic surfactant have the same hydrophobic entities namely $C_{12}$ alkyl chains and the same chemical group (ethylene oxide) as constituent of their hydrophilic parts. We showed that, at least up to a certain proportion of associative polymer in the mixed system, the hydrophobic cores remained identical to those of the pure surfactant micelles upon substitution of the surfactant by the polymer. We obtained evidence that the PEO chains linked preferentially the micelles. At low volume fraction this leads to an effective attractive interaction between the micelles and eventually to an associative phase separation.

To investigate further the effective interactions introduced by adsorbed associative polymers we adopt here other model systems. They consist of -oil in water- microemulsion droplets onto which hydrophobically modified hydrosoluble polymers are adsorbed. The associative polymers with two ends grafted or one end grafted with an hydrophobic chain consists of an alkyl chain -here $C_{12}H_{25}$ and of a water soluble poly(ethylene-oxide). The possibility for the associative polymer with two modified extremities (PEO-2m) to stick them in two different droplets can lead under appropriate conditions to a transient network forming a physical gel. The advantages of these model systems are threefold. The first advantage is that we can compare a situation where a transient network is indeed formed (upon addition of PEO-2m) to a very similar situation where no network can form (upon addition of a polymer modified at one end (PEO-m) the length of the PEO chain being here half that of PEO-2m). The second advantage is that we can monitor separately different parameters playing a role in the system, namely **i**: the radius of the droplets by adjusting the composition of the microemulsion, and their average distance depending on the volume fraction of the droplets. **ii:** the number of associative polymers added per droplet and **iii:** the length of the PEO chain. We can thus tailor the network adjusting the size and number density of knots and the number and length of the linking chains. This is a net advantage over similar networks formed in simple binary solutions of associative polymers where all these parameters are dependent on the chosen polymer and on the particular concentration used. The third advantage of our model systems is a practical one: we can easily obtain information on the form and size of the droplets and on the interactions introduced between them by studying the small angle neutron scattering patterns.

A similar strategy was developed previously by [23, 24]. They recently reported on the interactions introduced by associative polymers between microemulsion droplets. The studies were however limited to the case of associative polymers modified at both extremities.

The particular situations we plan to study have not been described theoretically but such a description would be closely related to that of the properties of telechelic polymers adsorbed onto surfaces [25,26] or in solutions [27]. We use these descriptions to picture the solutions in figure 1. If the added copolymer is a polymer modified at one end, it decorates the droplets as shown in figure 1A and the polymeric corona adds a repulsive contribution to the interaction between the droplets (i.e. an osmotic contribution to the repulsion , as long as the polymer is in a good solvent and the concentration is not too high). If the added polymer has two hydrophobic extremities they can adsorb onto the same droplet leading to a situation similar to fig 1A and/or they can adsorb onto two different droplets and, in that case, the polymers also introduce bridges between the droplets as shown in figure 1B. This can lead under appropriate condition to an effective attractive contribution to the interaction between the droplets and also to the formation of an interconnected network of droplets which must be responsible for the viscoelastic properties of such solutions.

The phase behaviour and the small angle neutron scattering (SANS) patterns of the samples are first described. The experimental results are then qualitatively discussed in terms of the effective interactions introduced by the PEO chains either bridging or decorating the microemulsion droplets.

**EXPERIMENTAL**

**Materials**:

Cetyl pyridinium chloride $[H_3C-(CH_2)_{15}]-C_5H_5N^+$ $Cl^-$ (CPCl) from Fluka is purified by successive recristallization in water and in acetone, octanol $[H_3C-(CH_2)_7]-OH$ and decane $[H_3C-(CH_2)_8 CH_3]$ from Fluka are used as received.

The poly (ethylene-oxide) have been hydrophobically modified and purified in the laboratory using the method described in[10,28]. The molecular weight of the starting products is determined by size-exclusion chromatography. After modification, the degree of substitution of the hydroxyl groups was determined by NMR using the method described in[29]. The degree of substitution is found to be equal or larger than 98%. The hydrophobically modified poly (ethylene-oxide) contains an isocyanate group between the alkyl chain $C_{12}H_{25}$ and the ethylene-oxide chain. We assume this isocyanate group belongs to the hydrophilic part of the copolymer. Poly (ethylene-oxide) PEO-2m is modified at both extremities and PEO-m is modified at one extremity only. PEO-m is exactly half the PEO-2m. cf Table 1.

All samples are prepared by weight in brine or deuterated brine. Brine-0.2M NaCl- is prepared with triply distilled water or deuterated water from S.D.S. used as received. The samples are characterized by the volume fraction $\Phi$ of hydrophobic cores [HC] which are composed of the parts of the components indicated above in brackets. All the parameters necessary to calculate Φ from the sample composition are summarized in Table 1 .

**Preparation of the microemulsion** :

The microemulsion [30] is here a thermodynamically stable dispersion in brine of oil droplets surrounded by a surfactant film. We seek for conditions where the microemulsion is close to but below the line of emulsification failure. The line of emulsification failure is the limit above which the microemulsion droplets are saturated with oil and coexist with excess oil. On this line the microemulsion droplets have a radius corresponding to the spontaneous curvature radius of the surfactant film [31] . Under such conditions it is now well established that the droplets of microemulsion are spheres of a well-defined radius [32] and that they can be diluted over a large concentration range [33,34] . First the spontaneous radius of curvature of the surfactant film, composed of a surfactant -CPCl- and a cosurfactant -octanol-, is adjusted by varying its composition ; the ratio in weight of octanol to CPCl is set equal to 0.25. Then decane is added up to the emulsification failure limit and then a value slightly below this limit is adopted: the ratio in weight of decane to CPCl + octanol is, in what follows, equal to 0.62. We do find that the microemulsion can be diluted over the range of ~1 to 20 weight %. In this range the microemulsion droplets are fairly monodisperse spheres as described below.

**Preparation of the microemulsion droplets plus hydrophobically modified polymers:**

The samples are prepared by weight. Their overall composition is determined so as to obtain a constant volume fraction $\Phi$ of the hydrophobic parts (HC) of the droplets (which consist of the hydrophobic parts of the microemulsion constituents plus the alkyl chains of the PEO-m or PEO-2m) while increasing progressively the number of adsorbed alkyl chains. This is achieved by replacing a small amount of CPCl + octanol by the appropriate amount of modified PEO. The solutions are characterized by $\Phi$ and by the number of $C_{12}$ chains per droplet $\underline{r}$. To calculate $\underline{r}$ we assume that the radius of the spherical HC does not change with increasing substitution of the surfactant by the copolymers; we show below that this assumption stands up to $\underline{r} \sim 40$. It is important to stress here that, with the present choice of PEO-m and PEO-2m, a sample prepared with one or the other and with the same value of $\underline{r}$ contains the same number of $C_{12}$ chains per hydrophobic core but also the same total amount of EO units.

**Observation of the phase behaviour of the samples**

The samples prepared as described above are thoroughly shaken to insure homogenization and then kept at the temperature of observation, here T=20°C, in a thermostated water bath for several days before visual examination. When a phase separation is observed the samples are rehomogenized and set back to rest for a couple of days to confirm the observations.

**Small angle neutron scattering: SANS Measurements**

They have been performed at LLB-Saclay on the spectrometer PACE. The range of scattering vectors covered is $0.004 \text{ Å}^{-1} < q < 0.16 \text{ Å}^{-1}$. The temperature is T= 20°C. The scattering data are treated according to standard procedures. They are put on an absolute scale by using water as standard. And we obtain intensities in absolute units ($cm^{-1}$) with an accuracy better than 10%. In the samples studied here, the neutrons are scattered predominantly by the microemulsion droplets.

The small angle neutron scattering from colloidal solutions provides information on their structure [35,36]. If the colloidal aggregates can be assumed to be spherical or if, at least on average, the interaction potential between them has spherical symmetry, one can write the scattered intensity I (cm$^{-1}$) in the form:

$$I(q) = \Phi v (\Delta \rho)^2 P(q) S(q) = A P(q) S(q)$$
$$\text{with } A = \Phi v (\Delta \rho)^2 \qquad [1]$$

where q ( Å$^{-1}$ ) is the scattering vector; $\Phi$ is the volume fraction; v (cm$^3$) the dry volume of the aggregates and $\Delta\rho$ (cm$^{-2}$) the difference in the scattering length density of the aggregates and of the solvent. P(q) is the form factor of the colloidal aggregates and P(q->0)=1. S(q) is the structure factor which reflects interactions between the aggregates.

For monodisperse spheres of radius R :

$$P(q) = \left[\frac{3[(\sin qR) - qR \cos qR]}{(qR)^3}\right]^2 \qquad [2]$$

At high q values, the structure factor S(q->∞) = 1, the scattered intensity is therefore only related to the form factor of the colloidal aggregates. In this limit, we will use the Porod representation : $q^4 I(q)$ as a function of q which amplifies the oscillations of P(q). If P(q) is given by [2] then, in the Porod representation, the first and second maximum and the first minimum are such that:

$$q_{1\max} R = \frac{\pi\sqrt{3}}{2}, \quad q_{2\max} R = 2\pi \quad \text{and} \quad q_{\min} R = \frac{3\pi}{2} \qquad [3]$$

In order to take into account the possible polydispersity of the microemulsion droplets we assume that the size distribution is gaussian and write for $[A.P(q)]_{polydisperse}$ :

$$[A.P(q)]_{polydisperse} = \Phi(\Delta\rho)^2 \frac{4\pi}{3} \int \frac{R^3}{\sqrt{2\pi}\Delta R} \left[\frac{3[(\sin qR) - qR \cos qR]}{(qR)^3}\right]^2 e^{-\frac{(R-\bar{R})^2}{2(\Delta R)^2}} dR \qquad [4]$$

With $\bar{R}$ the mean radius and $\Delta R$ the standard deviation.

In the thermodynamic limit, i.e. $q \to 0$ the structure factor can be written [37]:

$$S(q \to 0) = \frac{k_B T}{\Phi v} \chi_T$$ where $\chi_T$, the isothermal compressibility, is defined as a function of the volume fraction $\Phi$ and of the osmotic pressure $\Pi$: $\chi_T = \left[\frac{1}{\Phi}\frac{\partial \Pi}{\partial \Phi}\right]^{-1}$

We can then write $$\chi_T = \frac{I(q \to 0)}{(\Delta\rho)^2 k_B T} \qquad [5]$$

At the spatial resolution of our scattering experiments ($\frac{\pi}{q_{max}} = \frac{\pi}{0.16} \sim 20\text{Å}$) the droplets are viewed as homogeneous particles consisting mainly of alkyl chains with one scattering length density : we "see" the hydrophobic cores [HC]. From the scattering lengths [38] and the densities of the components of the HC (cf table 1) the scattering length density is computed $\rho = -0.44 \; 10^{10}$ cm$^{-2}$ and in deuterated brine $\Delta\rho = 6.83 \; 10^{10}$ cm$^{-2}$ (see rel [1]). When some polymers are adsorbed onto the HC's, in the range explored here, we will assume that the contrast remains identical as the proportion of ethylene-oxide in D$_2$O remains low even in the corona surrounding the HC's. Let us remark that, as already stated, the above formalism strictly applies to monodisperse spherical aggregates interacting through a potential with spherical symmetry. Representing the attractive contribution introduced by bridging in terms of a spherically symmetrical potential is probably questionable. Nevertheless, we will use this formalism here, keeping in mind its obvious limitations, as a tool to compare the effective interaction introduced by PEO-m or PEO-2m.

**RESULTS**

**Characterization of the microemulsion:**

In figure 2 the SANS patterns for microemulsions with increasing volume fraction $\Phi$ normalized to $\Phi=1$ are shown together with the spectra computed for spherical aggregates

Filali et al                                                                 10

interacting through a potential which is the sum of a repulsive and an attractive contribution (see below).

We first note that the oscillations of the form factor amplified in the Porod representation of figure 2A are perfectly superimposed in the large q range which is a strong indication that the droplets remain constant in size and shape in the range of $\Phi$ explored. Assuming the droplets are spherical we can deduce their radius from the position of the maxima and minimum in the Porod representation (fig.2A) using the relationships in [3]. All three extrema yield the same R value which confirm our assumption that the droplets are spherical; we obtain R = 62± 2 Å. The moderate damping of the oscillations must be traced back to size polydispersity or size fluctuations [39], they are correctly taken into account if we assume a gaussian distribution of radius for the microemulsion droplets with a mean radius $\overline{R}$ = 62Å and a standard deviation $\Delta R$ = 8 Å which indicates a small polydispersity. From the mean radius and the known composition of the samples we can calculate that each droplet of microemulsion is on the average composed of 830 CPCl, 542 octanol and 1537 decane and that its molar mass (including the polar heads) is roughly equal to $5.7 \, 10^5$ Dalton.

Looking at figure 2B the large drop of the intensity at small q's and the progressive appearance of a correlation bump, as $\Phi$ increases, are clear evidences that the microemulsion droplets experience an overall repulsive interaction. We compute the spectra shown as solid lines in fig.2A and 2B assuming the droplets to be spheres with R = 62Å interacting through a potential which is the sum of an attractive Van der Waals potential with an Hamaker constant appropriate for decane droplets interacting through water ($A_H$ = 1.1$k_B$T) and of a repulsive potential with a hard sphere contribution and a screened coulombic contribution for which the Debye length is determined through the small ions added ($\lambda_D$ = 7Å). To describe this effective potential the number density of macroions (=the microemulsion droplets) can be computed from $\Phi$ and the mean value of R so that there is only one adjustable parameter in the model namely the effective charge Z on each macroion. With this effective potential, the structure factor S(q) is calculated from the Ornstein-Zernike integral equation for the direct and total correlation function using Percus-Yevick closure relation via a program developed by Belloni [40]. Then the spectra can be calculated from [1] using the computed S(q) and [A. P(q)]$_{polydisperse}$ given by [4] and calculated with the estimated value of $\Delta\rho$ (see above) and $\bar{R}$ = 62Å and $\Delta R$= 8Å. An excellent quantitative agreement is achieved, as can be seen upon inspection of figure 2B, for the spectra at $\Phi$ = 1.4, 2.8, 7% with the same effective charge Z= 150 on each droplet. At $\Phi$ = 13.7% the position of the correlation bump is correctly computed but the drop of the intensity at q=0 is underestimated illustrating the limits of the model used. The effective charge Z can be related to the number of CPCl molecules in the surfactant film surrounding a droplet ~ 830 so that the ratio of dissociated surfactant ions to total surfactant ions is equal to ~ 0.2, as currently found in the case of ionic micelles in brine.

**Addition of unmodified PEO to the microemulsion**

In order to make sure that the PEO polymeric chains do not interact with the microemulsion droplets we compare, in figure 3, the SANS spectra of a microemulsion with $\Phi= 7\%$ to the SANS spectra of the same microemulsion with an amount of PEO (M=10000) corresponding to that added in a sample containing PEO-2m with <u>r</u>, the number of C12 chains per droplet,=12 . This illustrates that there is no interaction between the hydrophilic part of the polymer and the cationic surfactant in excellent agreement with previous studies [15], this is in contrast with comparable situations where the surfactant is anionic.

**Addition of modified PEO to the microemulsion**

*Do the microemulsion droplets remain identical ?*

We expect that adding PEO-m or PEO-2m to the microemulsion does not modify the hydrophobic droplets. This can be tested by comparing the SANS patterns in the Porod representation for the bare microemulsion and for the samples with PEO-m or PEO-2m added. Such a comparison is made in figure 4 for samples with $\Phi = 7\%$ and <u>r</u> = 8 ; the three patterns are perfectly superimposed in the large q range indicating no change in the hydrophobic droplets. Similar observations have been made for all samples with $0<\Phi<20\%$ and $0<$ <u>r</u> $<40$ . This provides strong evidence that the droplets remain identical. Furthermore this also help casting out the existence of aggregates of the associative polymers alone. The associative polymers in binary solutions are know to self-assemble forming small hydrophobic domains surrounded by the PEO polymeric chains, typical size for these domains have been found to be R~ 15-20Å [13] . Existence of such aggregates, coexisting with the microemulsion droplets, ought to show up in the SANS spectra; in the present experiments, even at the highest <u>r</u> and $\Phi$ values, no modification of the spectra in the high q-range could be detected. The assumption that the modified polymers associate with the microemulsion droplets rather than self-assembling separately was made here from the start considering the gain of entropy for the PEO chains to distribute themselves among large hydrophobic droplets rather than to squeeze around the small hydrophobic domains the polymers could eventually built up. This will be further confirmed in what follows by the effective interactions clearly introduced between the hydrophobic droplets upon addition of the polymers.

Having established that the droplets remain identical for samples with PEO-m or PEO-2m added in the range $0<\Phi<20\%$ and $0<\underline{r}<40$ we can now examine the interactions introduced by the presence of the PEO polymeric chains at the surface of the droplets. To do so we have two sets of experimental evidences namely the phase behaviour of both systems and the evolution of the SANS patterns in the low q range.

### *Phase Behaviour of the microemulsion plus PEO-m or PEO-2m*

As already stated the microemulsion remains a clear phase for volume fractions of the hydrophobic core between 1 and 30%. We investigate the evolution of the samples with PEO-m and PEO-2m added at T= 20°C

In figure 5 we sketch the phase evolution of the samples with $\underline{r}$ the number of C12 chains per hydrophobic core (plotted as vertical axis) and with the volume fraction $\Phi$ of hydrophobic core (plotted as horizontal axis). The situation is very different with one or the other copolymer. Upon addition of PEO-m (cf figure 5A) in the explored range of volume fraction, the samples remain clear monophasic solutions whatever the quantity of copolymer added up to $\underline{r}$ ~40. In contrast (cf figure 5B), for low volume fractions < ~ 10%, increasing the quantity of PEO-2m leads to a phase separation: a dilute phase coexists with a more concentrated phase. On the other hand at higher volume fractions the samples remain clear monophasic solutions when PEO-2m is added. In a similar system studied by Bagger-Jörgensen such a phase behaviour is also reported added [24]. This phase separation is an associative phase separation [41] brought about by an effective attractive interaction between the droplets. This interaction which must be due to the bridging of the droplets by PEO-2m is discussed further below.

With PEO-m a phase separation takes place at higher polymer concentration (r > 40) are added as indicated in figure 5A. This phase separation is different in nature to that observed with PEO-2m. The solution becomes turbid and a phase separation occurs with a thin oily phase floating above a clear solution. We will discuss this phase separation below; it is related to an emulsification failure with a fraction of the decane being rejected while the clear phase is a microemulsion of smaller drops with a larger curvature and less oil.

### *Comparison of the SANS patterns upon addition of PEO-m or PEO-2m*

In figure 6 the scattering patterns are represented for samples at fixed volume fraction of hydrophobic cores and an increasing number of copolymer added. At $\Phi = 7\%$ illustrating, in the low volume fraction range, the noticeable difference between samples with PEO-m or PEO-2m added. In figure 6A note the increasing position correlations induced by the decoration of the droplets by PEO-m as indicated by the peak which becomes increasingly more pronounced as r increases from 0 to 12 And in figure 6B, where the ordinate scale is double with respect to that in 6B, two features are noteworthy, first a very important rise of the intensity scattered at very low q's as r increases from 0 to 8 ( r = 12 cannot be studied as the samples phase separate as discussed above) and second the indication that a particular distance between droplets is favoured giving rise to an increasing correlation peak less marked however than with PEO-m.At $\Phi = 13.7\ \%$ illustrating the behaviour in the high volume fraction range where addition of both polymers amplifies the correlation maximum and where the difference between PEO-m and PEO-2m vanishes progressively with increasing $\Phi$. In figure 6C and figure 6D the vertical scale is the same as in figure 6A ; note in figure 6D that a small rise of the intensity scattered at low q's is still noticeable for PEO-2m.

In figure 7 the scattering patterns for samples with a constant r and varying volume fraction F are displayed. To allow an easy comparison we plot the intensity normalized to unit volume fraction as a function of q. Note first that here again the vertical scale of B and D relative to samples with addition of PEO-2m are twice those of A and C the analogous plots for samples with addition of PEO-m. This underlines the very different evolution of the scattering patterns in the two series of samples.

For PEO-2m samples, an important rise of the scattered intensity at low q is the signature of the effective attractive interaction (which must be due to the bridging) between droplets introduced by PEO-2m which is not observed in the samples with PEO-m. This attractive interaction is found to become less and less important when $\Phi$ increases above 7% and, in fact, the patterns for the two samples with $\Phi = 30\%$ and PEO-m or PEO-2m can be seen to be almost identical we will comment on this below. In PEO-m samples the depletion of the intensity at low q's is the signature of the repulsive interaction between droplets.

For both series of samples a broad correlation peak is observed which becomes sharper when $\Phi$ increases; this is the signature of increasing correlation between the droplets. The maximum of the peak is roughly at $q_{max} = 2\pi / d$ with $\underline{d}$ the average distance between nearby droplets. However here again the situation is very different in the two series of samples. In PEO-m (fig 7A and 7C) the maximum of the correlation peak moves to larger and larger q's as $\Phi$ increases as is expected for an ordering due to a regular repulsive interaction. In fig 8 we plot the evolution of $\underline{d}$ as a function of $\Phi$ for the samples with added PEO-m we note that it decreases with increasing $\Phi$ as in the microemulsion: $\underline{d} \sim \Phi^{-1/3}$. For the samples with added PEO-2m we note upon inspection of figure 7B and 7D that the maximum of the peak remains at the same position at the lowest $\Phi$'s and correlatively in figure 8 it can be seen that d remains constant at low volume fraction but its evolution becomes identical to that of the corresponding samples with PEO-m at higher Φ's .

**DISCUSSION**

From the above described experimental results we can draw a rather precise qualitative picture of the structure of the microemulsion droplets with addition of one or the other associative polymers. Let us recall that up to r ~ 40 the droplets are found to remain spheres of the same form -roughly spherical- and same average size R=62Å than in the bare microemulsion.

The PEO-m decorate these droplets introducing a repulsive contribution to the interaction, evidence of a local order is clearly given by the correlation peak: this order is very similar to that of the bare microemulsion but reinforced; this is clearly indicated by the average distances deduced from the position of the correlation peaks as shown in figure 8 and in table 2.

In the straightforward situation of the microemulsion decorated by the PEO chains of the PEO-m we already mentioned that a phase separation is observed when r becomes larger than 40: oil is ejected from the droplets, evidence from SANS that the droplets become smaller is shown in figure 9. While in the Porod representation it can be seen that the superposition of the oscillations of the form factor are perfectly superimposed for the bare microemulsion and samples with PEO-m r=8 or r=32 in figure 9A , these oscillations are moved towards higher q for the sample with r= 64 in figure 9B. From the position of the maximum and minimum we deduce, using relations [3] , R= 50Å.. When preparing the microemulsion we seeked for a situation close to the emulsification failure line, that is the surfactant film surrounding the decane droplets is close to its spontaneous curvature, the observation made here, of oil ejection while the droplets of microemulsion become smaller, points out to a change in the spontaneous curvature of the surfactant film which can easily be understood if we recall that the surfactant film is now decorated by PEO coils. Lateral repulsions of the PEO coils, decorating the surfactant film on the aqueous side, induce extra bending which increases with the area density of polymer. In other words there is a coupling between the elastic properties of the surfactant film and its decoration by the PEO. Assuming that pure decane is rejected that is the total area of the surfactant film remains constant we can evaluate the percentage of decane rejected by a simple calculation:

$\frac{\Delta \Phi}{\Phi} = \frac{\Delta R}{R} \sim 0.19$ in agreement with the quantity of decane floating over the sample.

We turn back to the main subject of this paper and to the samples with r<40. Comparison of the behaviour of samples with PEO-2m to that of samples with PEO-m points to an important difference between them. As already indicated the main difference one can imagine between the two series of samples is that PEO-2m can bridge the droplets (see fig 1). The observed differences are thus readily attributed to preferential bridging occurring when PEO-2m is added. In the range of low volume fractions ($\Phi < 10\%$) the addition of an attractive component to the interaction is clearly indicated by the rise of the intensity in the low q's range. How can we describe the samples ? In fact the attractive component can be pictured as due to the longer time the droplets spend close to one another because of the preferential bridging. The scattering patterns give us another clue namely that a preferential distance between the droplets exists in the samples as indicated by the existence of a correlation peak. This distance is independent of $\Phi$ (for $\Phi<10\%$) and of r as can be seen in figure 8 (see also table 2) it is also much smaller than the average distance calculated from $\Phi$, assuming short range simple cubic order, and observed in the bare or decorated microemulsion. It is tempting to deduce from this distance d the length of the PEO chain bridging two droplets we easily find ~100Å. From the relationship for the radius of gyration :

$R_g = 0.107 \pm 0.001 * M^{0.63 \pm 0.01}$ for a PEO chain of molar mass M [42, 43] we obtain $R_g = 35$Å and deduce for the average end-to-end distance of the polymer chain ~ 90 Å [44] in agreement with the value

Filali et al                                                                                                                    18

estimated above. We can then imagine the samples as a dispersion of temporary superaggregates of droplets linked by the PEO-2m bridges.

At this stage we must question: why is bridging favoured compared to decoration ? For the hydrophobic extremities the gain in energy is identical no matter whether their are in the same or in two different droplets as long as they experience the same apolar environment. When the droplets are far apart, the chains are too short to bridge them: a chain having one of its stickers in one given droplet is forced to loop so that its second sticker adsorbs onto the same droplet. When the droplets are close enough to one another (at a distance of order $R_g$), loop conformations are still allowed and, in addition, bridging conformations are now accessible. So the conformational entropy of the chain is larger when the droplets are at the right distance for bridging. This very simple argument was first proposed by Witten in ref [45] : assuming that the numbers of loop and bridge conformations are roughly equal, he derived the free energy change in bringing the droplets close to each other: -$k_B T$ ln(2) per telechelic chain. More refined calculations were reported [25,46,47] later on for the effective bridging interaction between flat surfaces in the different regimes (mushroom and brush) for the area density of telechelic chains: a neat attraction is again found but with a magnitude somewhat lower. The effect of the bridging attraction onto the phase behaviour of solutions of associating polymers are further analysed theoretically in [27] .

At higher volume fraction the droplets are on the average closer than the end-to-end distance of the polymer chain and bridging can occur without adding any attractive component to the interaction between droplets. On the contrary the bridging polymer chains will add to the repulsion in the same way as the decorating polymer chains resisting to interpenetration and longing to swell in brine. But the existence of bridges in the samples is still reflected macroscopically by their viscoelastic properties.

We have, indeed, reached our initial aim and built up a model system for transient networks. Its dynamical properties will be described in a forthcoming paper.

The interactions between droplets are reflected by the evolution of $\chi_T$ with the volume fraction of droplets and with the number of C12 chains per droplet. We evaluate the value of the osmotic compressibility in the following way. First we deduce $I(q->0)$ from the scattering patterns and then evaluate $\chi_T$ using relation [6] with $\Delta\rho = 6.83 \; 10^{10} \, \text{cm}^{-2}$ which gave an excellent agreement between the experimental and calculated scattering patterns of the bare microemulsion as described above. In figure 10, the curves of $\chi_T$ as a function of $\Phi$ reflect the features of the effective interactions introduced upon addition of PEO-m or PEO-2m to the bare microemulsion. Upon addition of PEO-m the compressibility decreases slightly indicating the reinforced repulsive interaction. But upon addition of PEO-2m we note a strong maximum in the curves; the volume fraction of the maximum, $\Phi_{max}$, is a value between 5 and 7% . This evolution is precursor to the associative phase separation observed in this system which is due to the effective attraction brought about by bridging. This phase separation (see figure 5) occurs for samples with $r > 12$ and with a critical volume fraction of ~ 6% where the osmotic compressibility is known to diverge, this value is in excellent agreement with the position of the maximum found here.

Acknowledgements.

One of us, R.A., is grateful to Michel Viguier and André Collet for their precious advices during the synthesis of PEO-m and PEO-2m.

We thank Loic Auvray and Didier Lairez for their help during the SANS experiments performed on line PACE at Laboratoire Léon Brillouin -CEA-CNRS.

We gratefully acknowledge the gift by Dr Belloni of the program he developed for the calculation of S(q) in terms of the interaction potential.

Table 1 Molar Mass and density of the components of the samples

| Component (abreviated in the text) | Molar Mass (dalton) | Molar Mass (Dalton) HC [a] | Density (g/cm$^3$) polar part | Density (g/cm$^3$) HC [a] |
|---|---|---|---|---|
| $D_2O$ | 20 | - | 1.105 | |
| $[H_3C-(CH_2)_{15}]-C_5H_5N^+ Cl^-$ (CPCl) | 339.5 | 225 | 1.656 | 0.83 |
| $[H_3C-(CH_2)_7]-OH$ (octanol) | 130 | 113 | 1.18 | 0.785 |
| $[H_3C-(CH_2)_8 CH_3]$ (decane) | 142 | 142 | - | 0.74 |
| $[CH_3-(CH_2)_{11}]-NH-CO-(0-CH_2-CH_2)_{227}$ $-O-(C0)-NH-[(CH_2)_{11} CH_3]$. (PEO-2m) | ~10 400 | 338 | 1.2 | 0.81 |
| $H_3C-(0-CH_2-CH_2)_{\sim 113} O-(C0)-NH-$ $[(CH_2)_{11} CH_3]$ (PEO-m) | ~ 5 200 | 169 | 1.2 | 0.81 |

a) HC = hydrophobic part of the component indicated in brackets in column 1

Table 2 $\chi_T$ and d extracted from the SANS patterns as explained in the text.

| Samples | | microemulsion | | microemulsion + PEO-m | | | | | | microemulsion + PEO-2m | | | |
|---|---|---|---|---|---|---|---|---|---|---|---|---|---|
| $\Phi\downarrow$ r$\rightarrow$ | (a) | 0 | | 4 | | 8 | | 12 | | 4 | | 8 | |
| | d(Å) | $\chi_T$ (c) | d(Å) | $\chi_T$ (c) | d(Å) | $\chi_T$(c) | d(Å) | $\chi_T$(c) | d(Å) | $\chi_T$ (c) | d(Å) | $\chi_T$(c) | d(Å) |
| 0.014 | 415 | 2.9 | (b)- | 2.3 | (b)- | 2.95 | (b)- | 2.6 | (b)- | 5.4 | (b)- | 8.5 | (b)- |
| 0.029 | 325 | 4.9 | (b)- | 4.6 | (b)- | 4 | 360 | 3.8 | 331 | 10.5 | 224 | 18.5 | 233 |
| 0.070 | 240 | 7.3 | 273 | 6.4 | 246 | 5.1 | 232 | 4.8 | 233 | 14.5 | 216 | 22.2 | 233 |
| 0.137 | 195 | 4.4 | 177 | 4.6 | 177 | 4.3 | 179 | 3.5 | 177 | 7 | 185 | 8.6 | 185 |
| 0.265 | 155 | 4.5 | 146 | 4.5 | 149 | 3.8 | 146 | 3.6 | 146 | 4.1 | 149 | 3.8 | 146 |

(a) calculated value of d the average distance between droplets assuming they are on a cubic network. From the volume fraction $\Phi$ and assuming the radius of the droplet R=62Å

(b) d is not measurable (no discernible correlation peak.)
(c) the osmotic compressibility, deduced from I(q->0) as explained in the text, is given in $10^{-6}$ Pascal$^{-1}$

**FIGURE CAPTIONS**

Figure 1: Illustration of decoration and bridging: the droplets of decane are surrounded by a monolayer of surfactant and cosurfactant and the poly(ethylene oxide) chains attached onto the cores by one or two stickers (a sticker= a $C_{12}$ alkyl chain) can decorate the droplets as shown in **1A** or, if and only if the copolymer is modified at both extremities, it can also bridge two droplets as shown in **1B**.

Figure 2 SANS Patterns for the microemulsion droplets as a function of the volume fraction $\Phi$ of hydrophobic cores: $\Phi$= 1.45%: $\Delta$; $\Phi$=2.9%: ; $\Phi$= 7 % O ;$\Phi$= 13.8 % +; the solid lines are the spectra computed as explained in the text. **2A**: the Porod representation of the spectra normalized to $\Phi$=1 illustrating, within experimental errors, the superimposition of the form factor in the high q range. **2B:** I(q) normalized to $\Phi$=1 (see text).

Figure 3 Comparison of the SANS Patterns for the microemulsion at $\Phi$ = 7% without :+ and with PEO (M=10 000): O.

Figure 4 The microemulsion droplets remain identical when PEO-m or PEO-2m is added: SANS patterns in the Porod representation for the bare microemulsion droplets: O, with PEO-m r= 8: and with PEO-2m r = 8: $\Delta$ in the three solutions the volume fraction of HC $\Phi$ is 7% .

Figure 5 Phase behaviour of the microemulsion upon addition of PEO-m: **5A** or PEO-2m : **5B**. The bare microemulsion is prepared here in water plus 0.2M NaCl,the ratio in weight of octanol to CPCl is kept identical (0.25) to the value adopted in deuterated brine but in that case the ratio in weight of decane to CPCl+ octanol must be slightly modified (=0.56) compared to the value adopted in deuterated brine in order to be in the same conditions with respect to the emulsification failure line.

Figure 6 Evolution with r, the number of C12 chains per droplet, of the effective interaction introduced upon addition of **6A** and **6C** PEO-m or **6B** and **6D** PEO-2m, SANS patterns for the bare microemulsion droplets :O, for r= 4 : and for r=8: $\Delta$.and for PEO-m only r=12: + Note that in 6B the ordinate scale is multiplied by 2. The volume fraction of hydrophobic core is $\Phi$ = 7% for **6A** and **6B** and $\Phi$ = 13.7% in **6C** and **6D**

Figure 7 Evolution with $\Phi$, the volume fraction of hydrophobic core, of the effective interaction introduced upon addition of **7A:** PEO-m or **7B:** PEO-2m, SANS patterns for $\Phi$ = 7% : O; $\Phi$ = 2.8 % : and $\Phi$ = 1.4 % :$\Delta$ and of **7C:** PEO-m or **7D**: PEO-2m, SANS patterns for $\Phi$ = 7% : O ; $\Phi$ = 13.7 % : and $\Phi$ = 26.5 %:$\Delta$. r, the number of C12 chains per droplet, is in all cases = 8

Figure 8 Position of the average distance between droplets (as deduced from $q_{max}$ the abscissa of the maximum of the correlation peak) as a function of $\Phi$, the volume fraction of hydrophobic core, in the spectra for the bare microemulsion: O, with added PEO-m r=4: $\Delta$ ; r=8: , r=12 :$\nabla$ and with added PEO-2m r= 4 : x and r=8 :+. For comparison calculated d (see text) is shown ♦ . The lines are just guides to the eyes.

Figure 9 Evolution of the size of droplets with r: SANS patterns in the Porod representation for solutions with, the volume fraction of hydrophobic core, $\Phi$ = 2.8 %. in the bare microemulsion droplets:O in 9A and 9B; **9A:** the droplets remain identical when decorated by PEO-m with r= 8 : $\Delta$ and r = 32 : and **9B:** the droplets become smaller when decorated with PEO-m with r=64 : full circle. (see text for discussion).

Figure 10  $\chi_T$, the osmotic compressibility, is plotted as a function of $\Phi$ for the bare microemulsion :O , with PEO-2m r=4: $\Delta$ and r=8:   ., with PEO-m r=4: + ; r=8: X and r=12 :$\nabla$.

**REFERENCES**

(1) TANAKA, F.; EDWARDS, S. F. *Macromolecules* **1992**, *25*, 1516-1523.

(2) *"Polymers in aqueous media, performance through association"*; GLASS, J. E., Ed., 1989; Vol. **223**.

(3) ODENWALD, M.; EICKE, E. F.; MEIER, W. *Macromolecules* **1995**.

(4) FRANÇOIS, J.; MAITRE, S.; RAWISO, M.; SARAZIN, D.; BEINERT, G.; ISEL, F. *Colloids & Surfaces A: Physicochem.Eng.Aspects* **1996**, *112*, 251.

(5) RASPAUD, E.; LAIREZ, D.; ADAM, M.; CARTON, J. P. *Macromolecules* **1996**, *29*, 1269-1277.

(6) SERERO, Y.; AZNAR, R.; PORTE, G.; BERRET, J. F.; CALVET, D.; COLLET, A.; VIGUIER, M. *Phys.Rev.Lett* **1998**, *81*, 5584-5587.

(7) CHAR, K.; FRANK, C. W.; GAST, A. P.; TANG, W. T. *Macromolecules* **1987**, *20*, 1833.

(8) *"Polymers as rheology modifiers"*; SCHULZ, D. N.; GLASS, J. E., Eds., 1991; Vol. **462**.

(9) KARUSEMA, J. E.; GLASS, J. E. *Prog.Org.Coatings* **1989**, *17*, 301.

(10) KACZMARSKI, J. P.; GLASS, J. E. *Macromolecules* **1993**, *26*, 5149.

(11) FRANÇOIS, J. *Prog.Org.Coatings* **1994**, *24*, 67.

(12) LUNDBERG, D. J.; BROWN, R. G.; GLASS, J. E.; ELEY, R. *Langmuir* **1994**, *10*, 3027.

(13) ALAMI, E.; RAWISO, M.; ISEL, F.; BEINERT, G.; BINANA-LIMBELE, W.; FRANçOIS, J. Model hydrophobically end-capped poly(ethylene oxide) in water. In *Adv.Chem.Ser* *""Hydrophilic polymers, performance with environmental acceptance"*; J.E.Glass, Ed., 1996; Vol. **248**; pp 343.

(14) CHASSENIEUX, C.; NICOLAI, T.; DURAND, D. *Macromolecules* **1997**, *30*, 4952-4958.

(15) BINANA-LIMBELE, W.; CLOUET, F.; FRANÇOIS, J. *Colloid Polym Sci.* **1993**, *271*, 748.

(16) ABRAHMSEN-ALAMI, S.; ALAMI, E.; FRANÇOIS, J. *J.Colloid Interf.Sci* **1996**, *179*, 20.

(17) ALAMI, E.; ALMGREN, M.; BROWN, W.; FRANCOIS, J. *Macromolecules* **1996**, *29*, 2229.

(18) ALAMI, E.; ALMGREN, M.; BROWN, W.; J., F. *Macromolecules* **1996**, *29*, 2229-2243.

(19) PERSSON, K.; WANG, G.; OLOFSSON, G. *J.Chem.Soc. Faraday Trans* **1994**, *90*, 3555.

(20) PERSSON, K.; BALES, B. L. *J.Chem.Soc.Faraday Trans.* **1995**, *91*, 2863.

(21) APPELL, J.; PORTE, G.; RAWISO, M. *Langmuir* **1998**, *14*, 4409-4414.

(22) BAGGER-JÖRGENSEN, H. Polymer effects on microemulsions and lamellar phases. PhD, University of Lund, 1997.

(23) GRADZIELSKI, M.; RAUSCHER, A.; HOFFMANN, H. *J.Phys. IV colloque C1* **1993**, *3*, 65.
(24) BAGGER-JÖRGENSEN, H.; COPPOLA, L.; THURESSON, K.; OLSSON, U.; MORTENSEN, K. *Langmuir* **1997**, *13*, 4204 - 4218.
(25) MILNER, S. T.; WITTEN, T. A. *Macromolecules* **1992**, *25*, 5495.
(26) SANJAY MISRA, W. L.; MAI NGUYEN-MISRA, W. L.; MATTICE, W. L. *Macromolecules* **1994**, *27*, 5037-5042.
(27) SEMENOV, A. N.; JOANNY, J. F.; KOKHLOV, A. R. *Macromolecules* **1995**, *28*, 1066.
(28) VOROBYOVA, O.; YEKTA, A.; WINNIK, A.; LAU, W. *Macromolecules* **1998**, *31*, 8998-9007.
(29) HARTMANN, P.; VIGUIER, M.; COLLET, A.; CALVET, D. *J. Fluorine Chemistry* **1998**, *in press*.
(30) SCHULMAN, J. H.; STOECKENIUS, W.; L.M., P. *J.Phys. Chem.* **1959**, *63*, 1677.
(31) TURKEVITCH, L. A.; SAFRAN, S. A.; PINCUS, P. A. . In *Surfactants in solution*; Mittal, K. L., Bothorel, P., Eds.; Plenum Press: New York, **1986**; Vol. 6; pp 1177.
(32) SAFRAN, S. A. *Phys.Rev. A* **1991**, *43*, 2903.
(33) CAZABAT, A. M. Probes of microemulsion structure. In *Physics of amphiphiles. Micelles, vesicles and microemulsions.*; DEGIORGIO, V., CORTI, M., Eds.; North Holland Pub.:, **1985**; pp 723-753.
(34) LEAVER, M. S.; OLSSON, U.; WENNERSTRÖM, H.; STREY, R. *J.Phys. II France* **1994**, *4*, 515-531.
(35) CABANE, B. . In *"Surfactant solutions: new methods of investigation"*; R.Zana, Ed.; M.Dekker Pub:, **1987**; pp 57.
(36) *Neutron, X-ray and Light scattering: introduction to an investigation tool for colloidal and polymeric systems*; LINDNER, P.; ZEMB, T., Eds.; North Holland: Amsterdam NL, 1991.
(37) HANSEN, J. P.; McDONALD, I. R. *Theory of Simple Liquids*; Academic Press: London GB, **1990**.
(38) Neutron-scattering-lengths. *Neutrons News* **1992**, *3*, 29-37.
(39) HAYTER, J. B. . In *Physics of amphiphiles: micelles, vesicles and microemulsions*; DEGIORGIO, V., CORTI, M Eds.; North Holland: Amsterdam NL, **1985**; pp 59.
(40) PEYRE, V.; SPALLA, O.; BELLONI, L.; NABAVI, M. *J.Colloid Interface Sci.* **1997**, *187*, 184.
(41) PICULELL, L.; LINDMAN, B. *Adv Colloid and Interface Sci* **1992**, *41*, 149.
(42) CABANE, B.; DUPLESSIX, R. *J.Phys. (France)* **1982**, *43*, 1529.
(43) BENKHIRA, A.; FRANTA, E.; RAWISO, M.; FRANçOIS, J. *Macromolecules* **1994**, *27*, 3963.
(44) de GENNES, P. G. *Scaling concepts in polymer physics*; Cornell Univ Press: Ithaca, USA; London, GB, **1979**.
(45) WITTEN, T. A. *J.Phys. (Paris)* **1988**, *51*, 1055.

(46)   LIGOURE, C.; LEIBLER, L.; RUBINSTEIN, M. *unpublished calculations, private communication* .

(47)   JOHNER, A.; JOANNY, J. F. *Euro.Phys. Lett.* **1991**, *15*, 265.

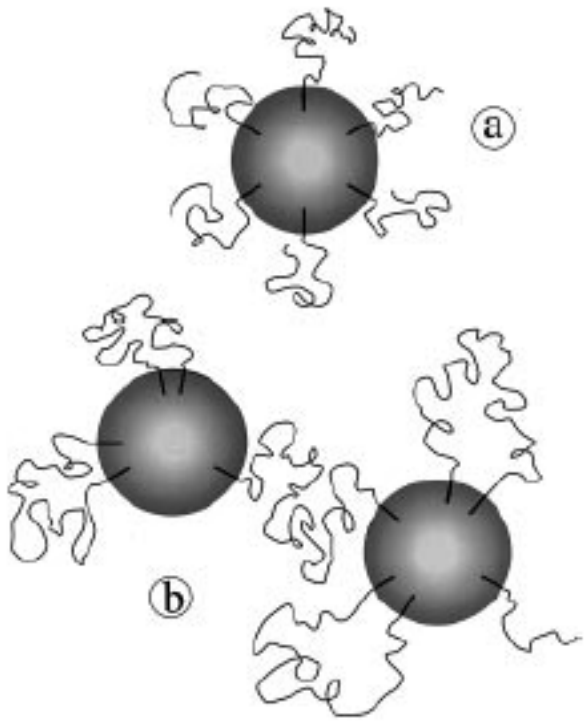

Filali et al  Figure 1

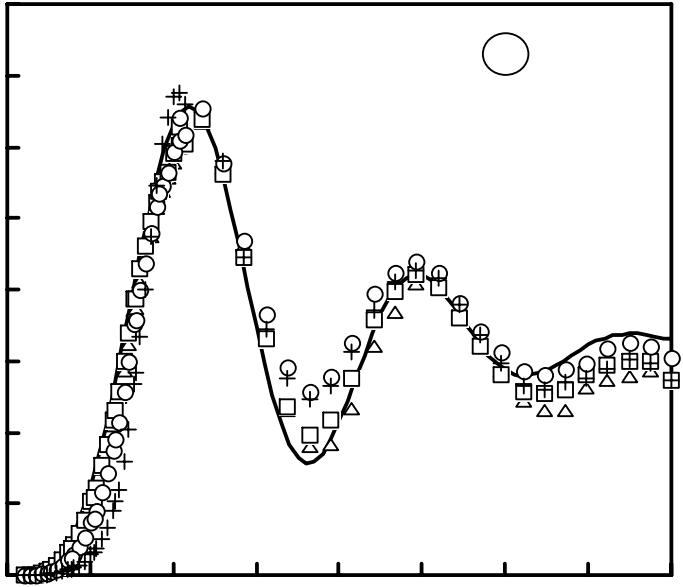

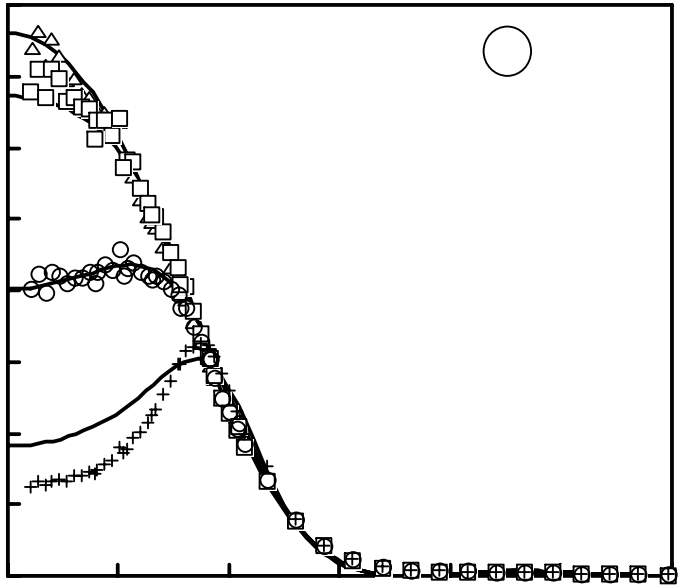

Filali et al Figure 2

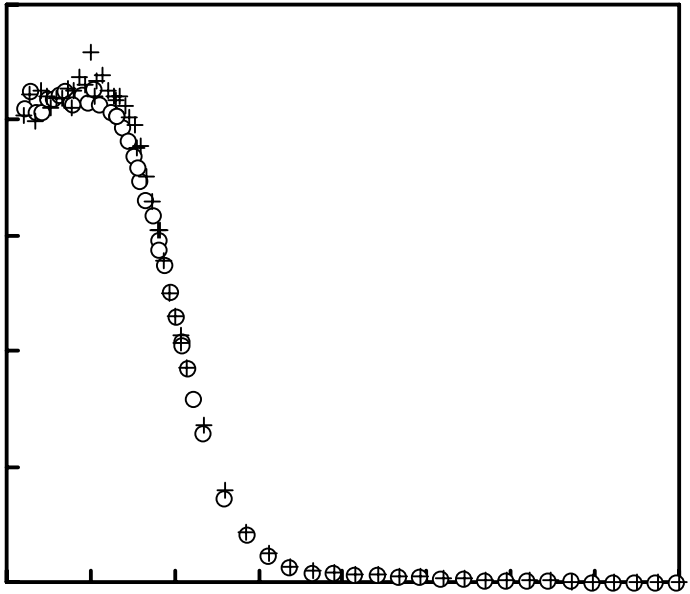

Filali et al Figure 3

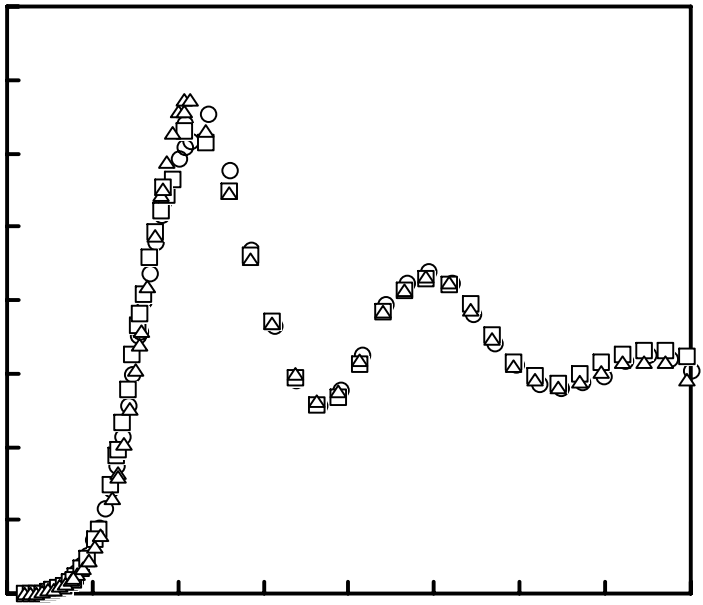

Filali et al Figure 4

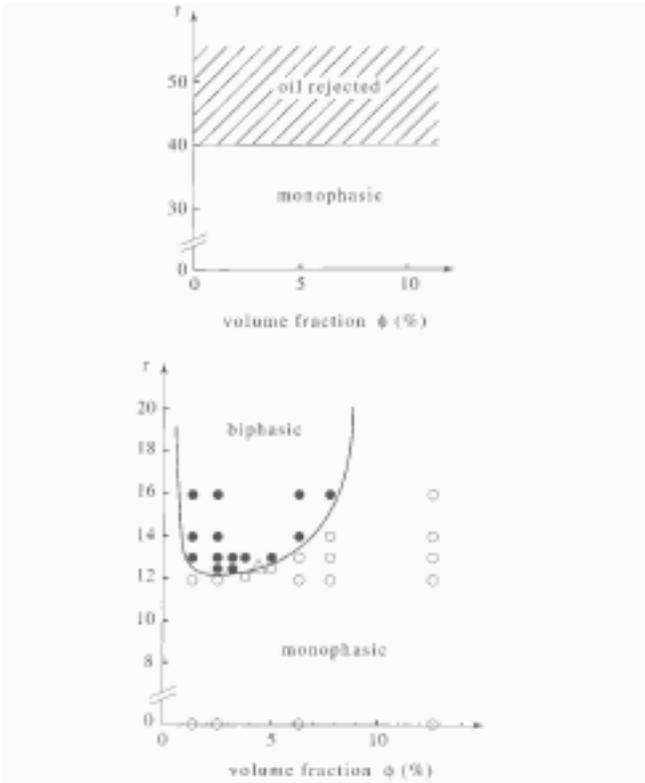

Filali et al Figure 5

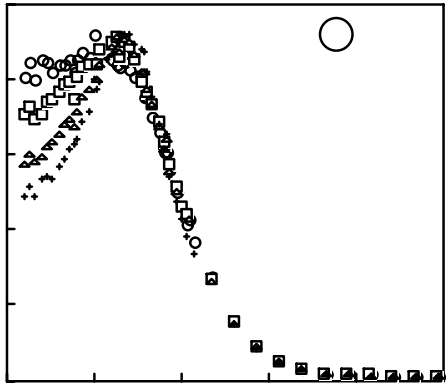 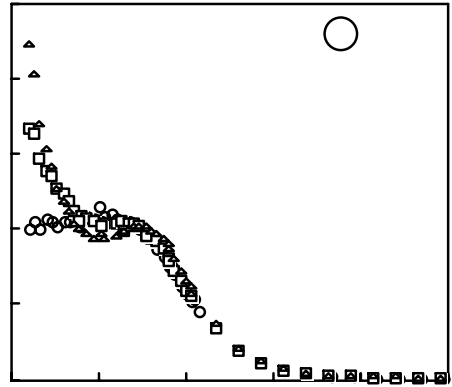

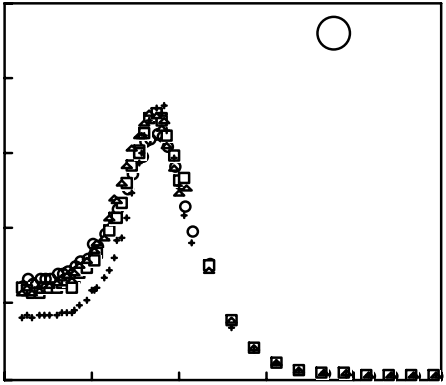 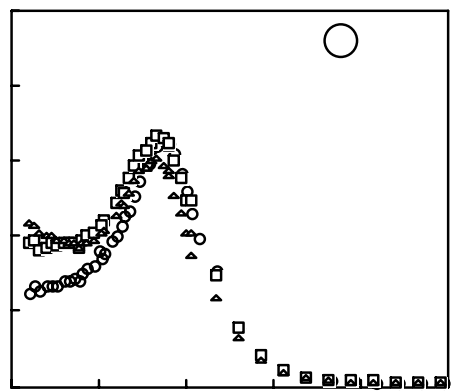

Filali et al Figure 6

Filali et al    33

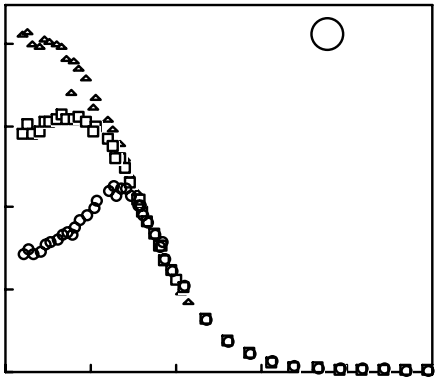
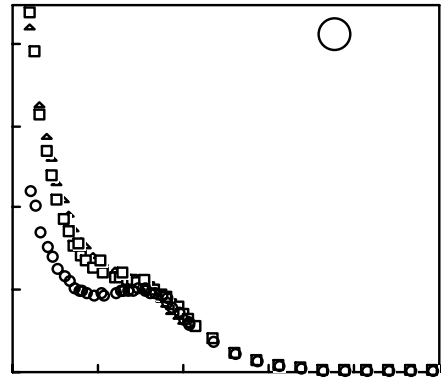
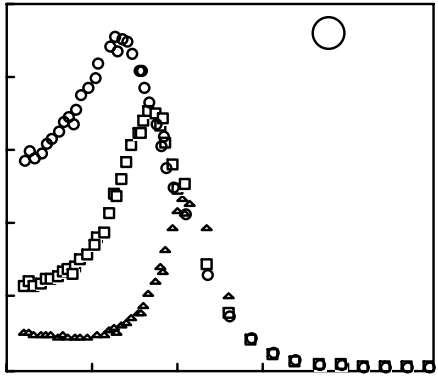
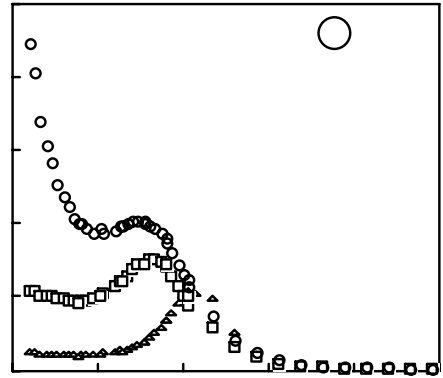

Filali et al Figure 7

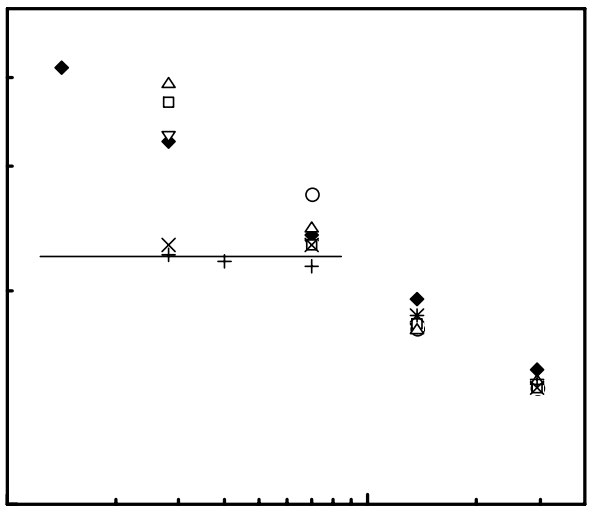

Filali et al Figure 8

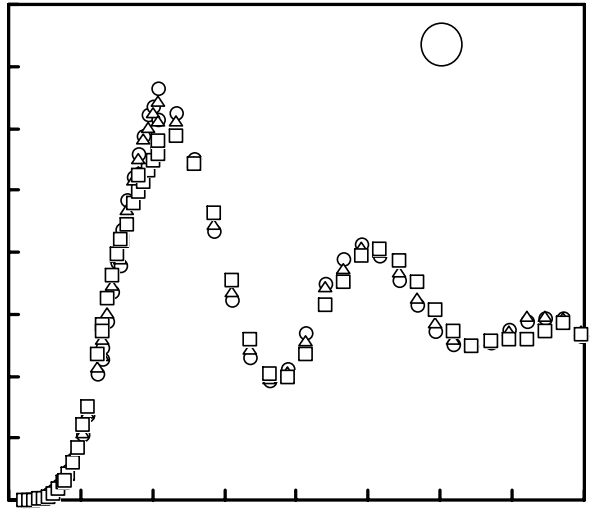

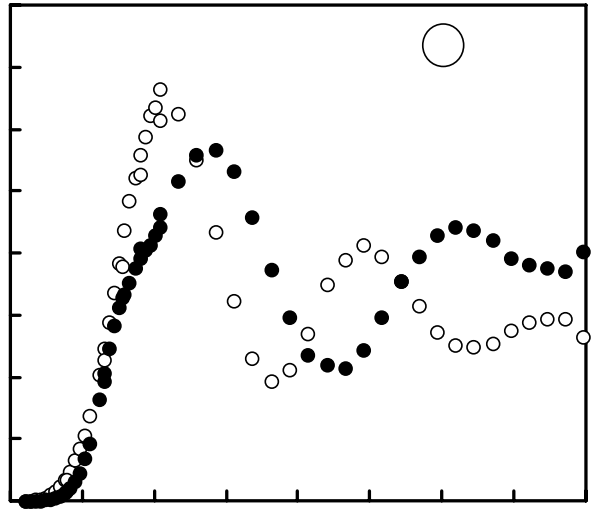

Filali et al Figure 9

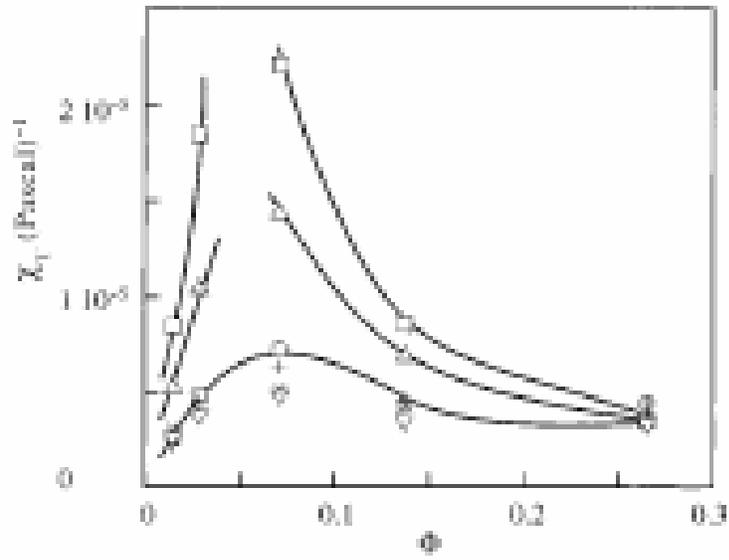

Filali et al Figure 10